\documentclass[graybox]{svmult}

\usepackage[sort&compress,numbers,square]{natbib}
\usepackage{mathptmx}       
\usepackage{helvet}         
\usepackage{courier}        
\usepackage{type1cm}        
\usepackage{makeidx}         
\usepackage{graphicx}        
\graphicspath{{Figures/}}
\usepackage{multicol}        
\usepackage[bottom]{footmisc}

\usepackage{url} 
\usepackage[colorlinks]{hyperref}

\usepackage[authormarkup=none]{changes}
\definechangesauthor[name={Stefan}, color=purple]{n}
\definechangesauthor[name={Stella}, color=blue]{s}
\definechangesauthor[name={Valerio}, color=red]{v}
\definechangesauthor[name={Alex}, color=green]{a}

\makeindex             

\newcommand{\la}{\langle}
\newcommand{\ra}{\rangle}

\newcommand{\mean}[1]{\ensuremath{\left\langle #1 \right\rangle}}

\newcommand{\cL}{\mathcal{L}}
\newcommand{\cD}{\mathcal{D}}
\newcommand{\cE}{\mathit{E}}
\newcommand{\cW}{\mathit{W}}
\newcommand{\cQ}{\mathit{Q}}
\newcommand{\cS}{\mathit{S}}

\newcommand{\da}{\dagger}

\newcommand{\Op}[1]{\hat{#1}}
\newcommand{\ophi}{\Op{\varphi}}

\newcommand{\osigma}{\Op{\sigma}}
\newcommand{\oL}{\Op{L}}
\newcommand{\oH}{\Op{H}}
\newcommand{\oA}{\Op{A}}
\newcommand{\oU}{\Op{U}}

\newcommand{\tr}{\ensuremath{{\rm tr}}}

\newcommand{\diff}{\mathrm{d}}
\newcommand{\curly}[1]{\ensuremath{\left\{#1\right\}}}

\newcommand{\eqref}[1]{(\ref{#1})}

\newcommand{\chtpi}{{\color[rgb]{0.4,0.2,0.9} \sc Qbook:Ch.1}} 
 
\newcommand{\chqfs}{{\color[rgb]{0.4,0.2,0.9} \sc Qbook:Ch.3}}

\newcommand{\chergo}{{\color[rgb]{0.4,0.2,0.9} \sc Qbook:Ch.8}}

\begin{document}

\title*{Quantum Rotor Engines}
\author{Stella Seah, Stefan Nimmrichter, Alexandre Roulet, Valerio Scarani}
\institute{Stella Seah \at Department of Physics, National University of Singapore, 2 Science Drive 3, Singapore 117542, Singapore, \email{stellaseah@u.nus.edu}
\and Stefan Nimmrichter \at Centre for Quantum Technologies, National University of Singapore, 3 Science Drive 2, Singapore 117543, Singapore, \email{cqtsn@nus.edu.sg}
\and Alexandre Roulet \at Department of Physics, University of Basel, Klingelbergstrasse 82, CH-4056 Basel, Switzerland, \email{alexandre.roulet@unibas.ch}
\and Valerio Scarani \at Department of Physics, National University of Singapore, 2 Science Drive 3, Singapore 117542, Singapore; and Centre for Quantum Technologies, National University of Singapore, 3 Science Drive 2, Singapore 117543, Singapore, \email{physv@nus.edu.sg}}
%
%
\maketitle


\abstract{This chapter presents autonomous quantum engines that generate work in the form of directed motion for a rotor. We first formulate a prototypical clock-driven model in a time-dependent framework and demonstrate how it can be translated into an autonomous engine with the introduction of a planar rotor degree of freedom. The rotor plays both the roles of internal engine clock and of work repository. Using the example of a single-qubit piston engine, the thermodynamic performance is then reviewed. We evaluate the extractable work in terms of ergotropy, the kinetic energy associated to net directed rotation, as well as the intrinsic work based on the exerted torque under autonomous operation; and we compare them with the actual energy output to an external dissipative load. The chapter closes with a quantum-classical comparison of the engine's dynamics. For the single-qubit piston example, we propose two alternative representations of the qubit in an entirely classical framework: (i) a coin flip model and (ii) a classical magnet moment, showing subtle differences between the quantum and classical descriptions.}

\section{Introduction}\label{sec:intro}

It is not easy to guess whether the performance of a thermal machine would improve or deteriorate, were the machine operated in the quantum regime. Intrinsic noise will certainly be a nuisance, while coherence and entanglement may prove additional resources if properly harnessed. Designing a thermal machine with ``quantum supremacy'' would be a major breakthrough in quantum thermodynamics. This chapter focuses on rotor engines, a promising testbed for identifying genuine quantum features in thermal machines.

By \textit{engine}, we understand a thermal machine that operates between two thermal reservoirs, the hot and the cold bath, transforming heat from the hot bath into \textit{useful energy}. In fact, the paradigmatic use of an engine is to generate \textit{directed motion}. With this in mind, we consider engines in which the \textit{working medium} is coupled to a \textit{rotor}\cite{alex2017rotor,stella2018,alex2018}. Its rotation leads to an unambiguous definition of mechanical output (work) in terms of directed motion.

The rotor engine will be \textit{autonomous}, like the engine of a car: once started, the cycle is self-sustained by interplay of gears and shafts (the driver's pedal controls the effective temperature of the hot bath). An autonomous engine keeps accelerating in the absence of friction or load\cite{tonner2005,youssef2010,brunner2012virtual,gilz2013,mari2015quantum,kosloff2016flywheel,alex2017rotor,hardal2017,stella2018,alex2018}; when these are present, they determine the timing of the cycle in the steady state. By contrast, textbook studies of engine cycles tend to consider \textit{driven} engines\cite{kosloff1984,scully2003,rezek2006,alicki2014,zhang2014,uzdin2016,stella2018}. Their dynamics is simpler, since the timing of the cycle is fixed, determined by an external clock. However, claims on the efficiency of such engines are to be made with care, as the energy needed to operate the driving system must be taken into account.


That being said, a rotor and a clock are very similar. We use this similarity in Sect. \ref{sec:driving} to introduce rotor engines. We start off with a textbook example of a time-dependent driven system and demonstrate how its time dependence can be mapped to a rotor degree of freedom. This clarifies also the dual role of the rotor in the autonomous engine: it can be seen both as an internal engine clock and as a work repository. We then discuss various definitions of work, notably comparing intuitive ones based on the rotor's motion with axiomatic ones like the ergotropy. In Sect. \ref{sec:classical}, we compare the dynamics of the quantum model with the corresponding classical dynamics. Since throughout the paper our working mode will be a qubit, this comparison will introduce two classical spin models: (i) a coin-flip toy model, based on biased thermal coin flipping rate equations; and (ii) a classical magnetic moment with linear coupling to harmonic oscillator baths.

\section{From clock-driven engines to rotor engines}\label{sec:driving}
\begin{figure}
\centerline{\includegraphics[width=\textwidth]{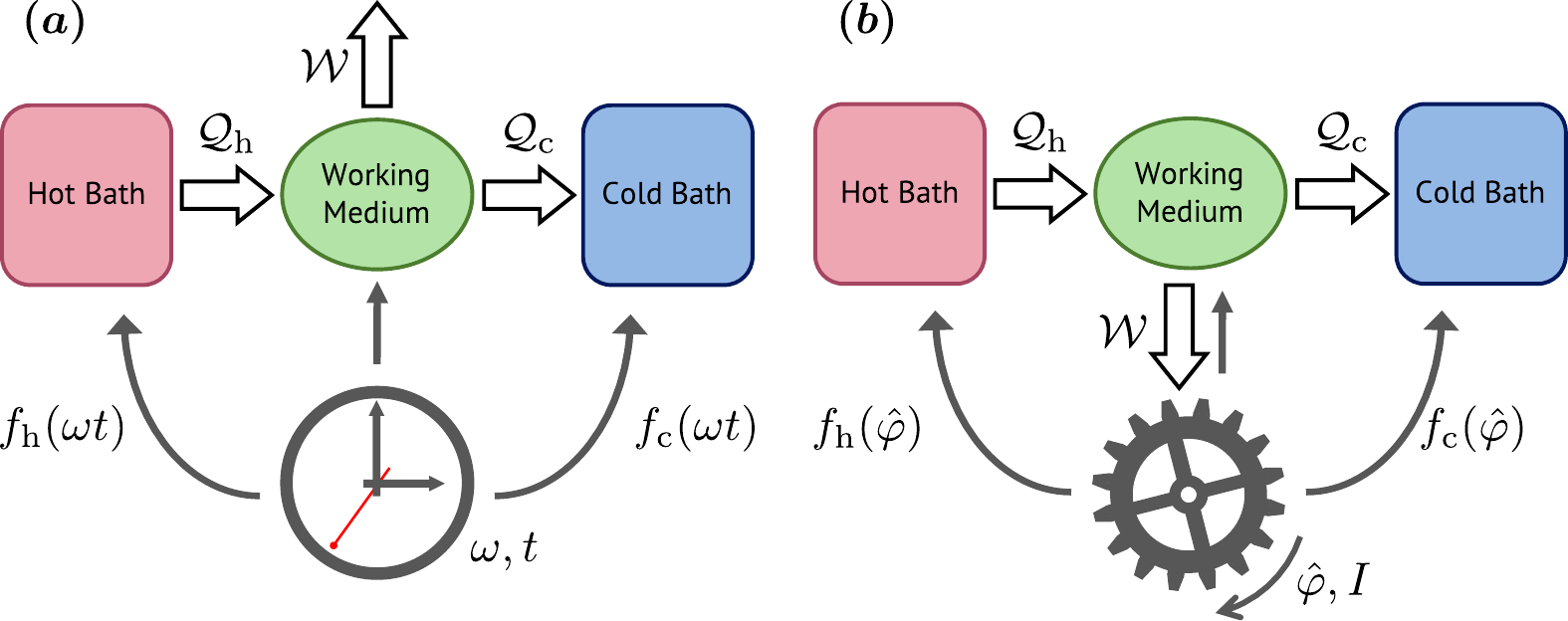}}
\caption{\label{fig:sketch} (a) Sketch of a clock-driven quantum engine with fixed cycle frequency $\omega$. The heat exchange between the engine's working medium and a hot and a cold reservoir is synchronized to the ticking clock pointer by virtue of periodic coupling functions $f_{\rm h,c} (\omega t)$, while a time-periodic modulation of the medium's energy provides the interface for work in- and output. (b) Autonomous engine version, where the clock is replaced by a quantum rotor with angle $\ophi$ and moment of inertia $I$. The fluctuating angular motion determines the cycle frequency, modulates the working medium energy and serves as a flywheel energy storage.}
\end{figure}

Consider a generic finite-time heat engine scheme comprised of a working medium and an external clock with characteristic frequency $\omega$ [Figure \ref{fig:sketch}(a)]. Its autonomous counterpart is obtained by replacing the clock with an embedded quantum rotor [Figure \ref{fig:sketch}(b)]. For the mathematical formulation in terms of a dynamical open quantum system model we will follow Alicki's derivation \cite{alicki1979quantum}.

\subsection{Mathematical model of a clock-driven engine}

As working medium, we choose a simple quantum system: a single or a few qubits, or harmonic modes with bare Hamiltonian $\oH_0$. The role of this medium is to perform and receive work $\cW$ and to mediate a heat transfer between a hot and a cold reservoir, $\cQ = \cQ_{\rm h} + \cQ_{\rm c}$. These reservoirs can be conventional thermal baths at different temperatures $T_{\rm h} > T_{\rm c}$, but we may also consider non-standard resources such as squeezed baths\cite{huang2012,obinana2014,rossnagel2014,niedenzu2016,klaers2017,wulfert2017} or continuous measurement processes that effectively inject entropy like a heat bath. 
Here, the coupling to the reservoirs shall be described by time-modulated Lindblad dissipators $\cL_{\rm h,c}^{\omega t}$ acting on the state $\rho$ of the working medium. 
Assuming the clock is sufficiently slow and the time modulations weak, one derives the master equation \cite{alicki1979quantum}
\begin{equation}\label{eq:MEtimeDep}
\partial_t \rho = -\frac{i}{\hbar} \left[ \oH_0 + \oH_{\rm int} (\omega t),\rho \right] + \cL_{\rm h}^{\omega t} \rho  + \cL_{\rm c}^{\omega t} \rho.
\end{equation}
What dissipators are used to model the reservoir interaction for a given working medium determines whether the master equation is valid and thermodynamically consistent. In the case of weak coupling to thermal baths, for example, consistency with the second law is achieved if each dissipator would by itself describe thermalization of the whole system, i.e.~drive the  state $\rho$ towards thermal equilibrium with the respective bath. However, when the working medium is a composite system, one often resorts to simpler dissipators that describe local thermalization of individual subsystems instead. Such an approximate treatment is only justified in the limit of weak intrinsic coupling between the subsystems \cite{rivas2010,levy2014,hofer2017,gonzalez2017}.

The function of the clock in \eqref{eq:MEtimeDep} is three-fold. First, it determines the duration $\tau = \frac{2\pi}{\omega}$ of an engine cycle. Second, it describes a modulation of the system energy by means of a periodic interaction Hamiltonian, $\oH_{\rm int} (\omega t) =\oH_{\rm int} (\omega t + 2\pi) $, and thereby provides the interface for work insertion and extraction. Third, it synchronizes the dissipative coupling between the working medium and the reservoirs in such a way that a net amount $\cW$ of work is generated over each cycle. Note that, instead of idealized sequences of clearly separate heat and work strokes, we consider here continuous engine cycles with harmonic modulations of the working medium and the thermal coupling strength, $\oH_{\rm int} (\omega t)$ and real-valued $f_{\rm h,c} (\omega t)$ to be specified below.

\subsubsection{Work output and energy input}

For the clock-driven engine, the work output is defined by separating the time-dependent modulation of the system Hamiltonian from the energy exchange with the reservoirs through the dissipators. The rate at which work is performed \emph{by} the system at some time $t$ is then given by the \emph{output} power
\begin{equation}\label{eq:workT}
\dot{\cW} (t) = -\tr \left\{ \rho(t) \partial_t \oH_{\rm int} (\omega t) \right\}.
\end{equation}
The energy \emph{input} from the reservoirs, on the other hand, can be expressed in terms of the rates
\begin{eqnarray} \label{eq:heatT}
\dot{\cQ} (t) &=& \dot{\cQ}_{\rm h} (t) + \dot{\cQ}_{\rm c} (t), \\
\dot{\cQ}_{\rm h,c} (t) &=& \tr \left\{ [\oH_0 + \oH_{\rm int} (\omega t)] \cL_{\rm h,c}^{\omega t} \rho(t) \right\} \approx \tr \left\{ \oH_0 \cL_{\rm h,c}^{\omega t} \rho(t) \right\}, \nonumber
\end{eqnarray}
where the approximation holds if we are in the weak-coupling limit where the modulations of the bare system energies due to the interaction is small. 
For thermal reservoirs, the input will be in the form of passive, disordered energy, i.e.~pure heat. But care must be taken in general, as the energy input of non-standard reservoirs could already contain a certain amount of useful work, see \chtpi.

The total energy change in the working medium over an engine cycle is 
\begin{equation}
\Delta E (t) = \int_t^{t+\tau} \diff t' \left[ \dot{\cQ}(t') - \dot{\cW} (t') \right] = \tr \left\{ [\oH_0 + \oH_{\rm int} (\omega t)] [\rho(t+\tau) - \rho(t)] \right\},
\end{equation}
which vanishes only in the quasi-static idealization where the system returns to its initial state after each cycle. For finite-time engines with finite-dimensional working media, the energy will typically grow over a number of periods before the engine reaches a steady limit cycle.

\subsubsection{Driven single-qubit piston engine} \label{sssec:QubitEngineTime}

As an instructive example that we will employ throughout the chapter, we consider a single qubit as working medium and the following model:
\begin{eqnarray} \label{eq:1QubitParamT}
\oH_0 &=& \hbar \omega_0 |e\ra \la e|, \quad \oH_{\rm int} (\varphi) = \hbar g \cos \varphi |e\ra \la e|, \\
f_{\rm h} (\varphi) &=& \frac{1+\sin\varphi}{2}, \quad f_{\rm c} (\varphi) = 1, \nonumber \\
\cL_{j}^{\varphi} \rho &=& \kappa f_j^2 (\varphi) \left\{ (\bar{n}_j + 1) \cD[\osigma_{-}]\rho + \bar{n}_j \cD[\osigma_{+}]\rho \right\}, \nonumber
\end{eqnarray}
with $\cD[\oA]\rho = \oA\rho\oA^\da - \{ \oA^\da \oA, \rho \}/2$, and with $\osigma_{+} = |e\ra\la g|$ and $\osigma_{-}=|g\ra\la e|$ the qubit raising and lowering operators.  
Here we introduced the pointer coordinate $\varphi = \omega t$ of the clock which we shall assume to point upwards to twelve o'clock at $\varphi=0$ and describing a clockwise rotation. The modulation $g \cos \varphi$ of the qubit frequency $\omega_0$ is then tied to the vertical position of the pointer (``piston''): the higher the piston, the higher the energy. Alternatively, it can be viewed as a constant pressure pushing down the piston when the qubit is excited, which corresponds to a time-dependent torque of magnitude $\hbar g$ responsible for the work exchange.

The thermal coupling to harmonic oscillator baths of temperatures $T_{\rm h}>T_{\rm c}$ and occupation numbers $\bar{n}_{\rm h,c} = 1/[\exp(\hbar \omega_0/k_B T_{\rm h,c})-1]$ is synchronized with the horizontal pointer position. Both baths are characterized by the same thermalization rate $\kappa$,
but while the cold one is in permanent contact to the qubit, the hot bath only couples appreciably if the clock points towards three o'clock ($\varphi=\pi/2$) or around. This exact model was not studied previously but is similar to those studied in Refs \cite{alex2017rotor,stella2018}. To ensure validity of the weak coupling master equation \eqref{eq:MEtimeDep}, we must further assume $g,\kappa,\omega \ll \omega_0, 1/\tau_{\rm h,c}$ with $\tau_{\rm h,c}$ the bath correlation times.

It is now intuitively clear how this piston engine generates net work as the clock performs one round trip. Starting from the upper-most position, the clockwise downward motion increases the hot bath coupling, which leads to a high average qubit excitation and downward pressure. Work is generated until the lower turning point at $\varphi=\pi$ is reached. The qubit is now predominantly coupled to the cold bath and therefore less excited on average, which results in less work being consumed in the subsequent upward motion of the pointer-piston.

\begin{figure}[t]
\sidecaption[t]
\includegraphics[width=7cm]{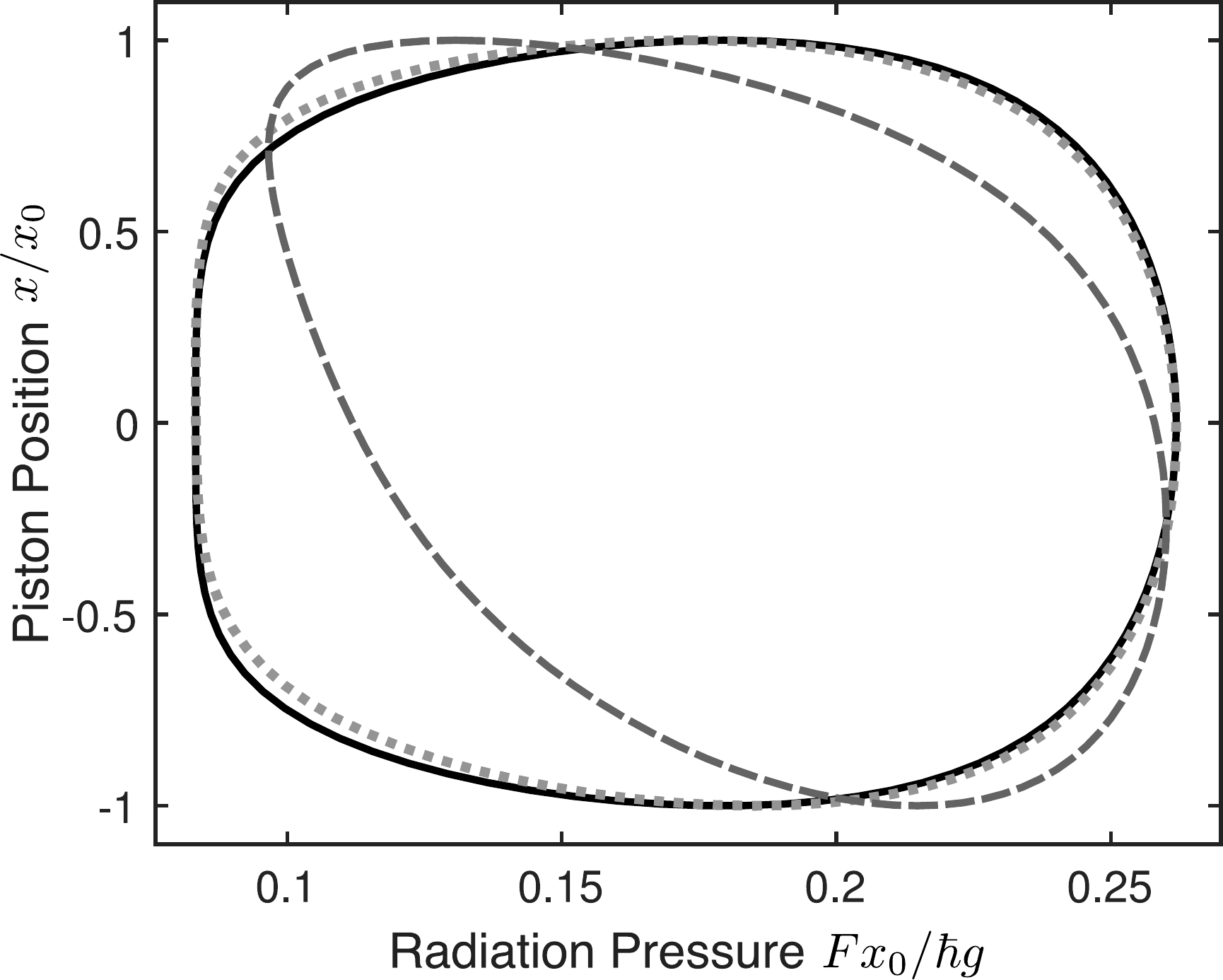}
\caption{Phase diagrams of driven single-qubit piston engine cycles. The horizontal and the vertical axis represent the average force by the qubit (pressure) and the vertical piston position (volume), respectively. We plot the ideal quasi-static cycle $\omega\ll \kappa$ (solid) and the finite-time limit cycles for $\omega/\kappa=0.1$ (dotted) and 1 (dashed). The parameters of the plot are $(\bar{n}_{\rm h},\bar{n}_{\rm c}) = (1,0.1)$ and $g=10\kappa$. For this example, the integral in \eqref{eq:Wqst} gives 0.31, the one in \eqref{eq:Qqst} gives 0.23.}
\label{fig:engineCycle}       
\end{figure}

Let us first analyse the quasi-static regime of slow rotation and fast thermalization, $\omega \ll \kappa$, where the qubit is approximately kept in thermal equilibrium. In the textbook case where the hot and the cold bath coupling occur in separate engine strokes ($f_{\rm h} (\varphi) f_{\rm c}(\varphi)=0$), the working qubit would thermalize to either mean excitations $p_{\rm h,c} = \bar{n}_{\rm h,c}/(2\bar{n}_{\rm h,c}+1)$, depending on the bath to which it is coupled. In the case of overlapping baths as we are considering here, at each time the qubit is in equilibrium at a $\varphi$-dependent effective temperature, with excitation probability
\begin{equation} \label{eq:pe_ideal}
p_e (\varphi) = \frac{\bar{n}_{\rm h} f_{\rm h}^2(\varphi)+\bar{n}_{\rm c} f_{\rm c}^2(\varphi)}{(2\bar{n}_{\rm h}+1) f_{\rm h}^2(\varphi)+(2\bar{n}_{\rm c} + 1) f_{\rm c}^2(\varphi)}.
\end{equation}
If we associate a vertical piston position $x = x_0 \cos \varphi$ to the clock angle, then the downward pressure of the excited qubit translates into an average force $F = \hbar g p_e (\varphi)/x_0$. This allows us to visualize the engine cycle in a phase diagram (Figure \ref{fig:engineCycle}). For finite values of $\omega/\kappa$, when \eqref{eq:pe_ideal} is not valid, $p_e (\varphi)$ is obtained from a numerical simulation of the dynamics, for which we used the QuTiP package \cite{johansson2013qutip}.

The enclosed area gives the \textit{net work output per cycle}. In the ideal quasi-static limit $\omega\ll \kappa$, this is given by \begin{equation} \label{eq:Wqst}
\cW_{\rm qst}^{\rm cyc} = \hbar g \int_0^{2\pi} \diff \varphi \, p_e (\varphi) \sin \varphi.
\end{equation} With growing $\omega$, the area of the limit cycles shrinks, as the thermalization lags behind the rotation of the clock pointer.

It is also instructive to look at the \textit{heat input per cycle} given by
\begin{equation} \label{eq:Qqst}
\cQ_{\rm h,qst}^{\rm cyc} = \hbar \omega_0 \frac{(2\bar{n}_{\rm h}+1) \kappa}{\omega} \int_0^{2\pi} \diff \varphi \, f_{\rm h}^2 (\varphi) \left[p_{\rm h} - p_e (\varphi) \right]
\end{equation} for $\omega\ll \kappa$. In the case of separate strokes, this would be $\hbar \omega_0 (p_{\rm h}-p_{\rm c})$ independent of $\omega$. Since the cold bath is always coupled, a slower cycle leads to an excess of heat input by a factor $\sim\kappa/\omega$.

\begin{figure}[t]
\includegraphics[width=\textwidth]{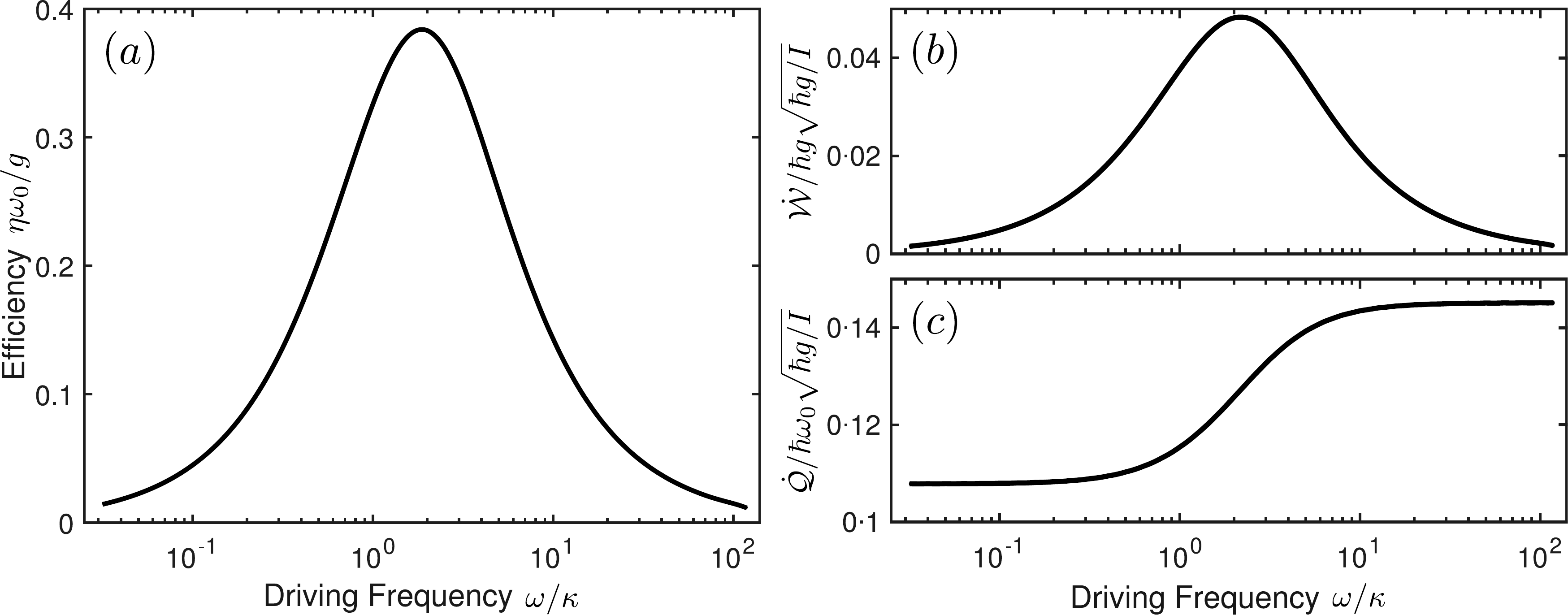}
\caption{Performance of engine as a function of the driving frequency $\omega$ for the same settings as Figure \ref{fig:engineCycle}. (a) Efficiency $\eta$ in units of $g/\omega_0$, where $g\ll\omega_0$. (b) and (c) Cycle-averaged work output $\dot{\cW}$ and heat input power $\dot{\cQ}$ respectively. We notice that the efficiency is mostly determined by $\dot{\cW}$, while $\dot{\cQ}$ is almost constant.}
\label{fig:effOutputTime}       
\end{figure}

Finally we turn to the \textit{efficiency per cycle} $\eta = \cW_{\rm qst}^{\rm cyc} / \cQ_{\rm h,qst}^{\rm cyc}$. In Figure \ref{fig:effOutputTime}(a), we plot it for varying engine frequencies $\omega$ at otherwise fixed rates $g,\kappa$ as before. The engine attains its maximum efficiency $\eta\approx 0.4 g/\omega_0\ll 1$ for $\omega \sim \kappa$ (i.e.~away from the quasi-static limit).

Panels (b) and (c) show the behavior of the cycle-averaged \emph{powers} of work output and heat input, respectively.
The efficiency is mainly determined by the behavior of the work output power, which deteriorates once the engine rotates faster than the qubit can thermalize. In this regime where $\omega\gg\kappa$, the per-cycle work output fails to reach the quasi-static optimal value \eqref{eq:Wqst}.

As for the heat input power, it assumes its quasi-static minimum \eqref{eq:Qqst} at low frequencies and increases slightly with $\omega/\kappa$, indicating that the heat leak from the hot to the cold bath worsens when thermalization lags behind.

A velocity-dependent efficiency and output power with a sweet spot of optimal performance is typical for classical piston engines and motors. In the driven engine just discussed, it depends entirely on the intrinsic reaction time $1/\kappa$ of the working medium. The dependence will be different once we replace the external time dependence that drives the engine by an autonomous rotor clock and introduce an actual load for steady-state work extraction.

\subsection{Reformulation as an autonomous rotor engine}\label{ssec:autoEngine}

Now we replace the external clock with a built-in engine clock. To this end, we replace the regular ticks at a fixed frequency $\omega$ with a planar rotor degree of freedom characterized by its moment of inertia $I$ and its canonical variables $\left(\ophi,\oL\right)$, as sketched in Figure \ref{fig:sketch}(b). The dynamical variable $\ophi$ of the rotor now serves as an internal clock that sets the engine cycle\footnote{To ensure a consistent quantum description of the rotor angle, the operator $\ophi$ will appear only in the form of strictly $2\pi$-periodic functions.}. Concurrently, the rotor would act as an integrated ``flywheel" that captures the work injected by the thermally driven working medium in the form of kinetic energy, $\hat{L}^2/2I$. The master equation for the autonomous engine, replacing the previous \eqref{eq:MEtimeDep}, is now given by 
\begin{eqnarray}\label{eq:MEauto}
\partial_t \rho &=& -\frac{i}{\hbar} \left[ \oH_0 + \oH_{\rm int}  (\ophi)+ \frac{\oL^2}{2I},\rho \right] + \cL_{\rm h}^{\ophi} \rho  + \cL_{\rm c}^{\ophi} \rho,\\
\cL_{j}^{\ophi} \rho &=& \kappa (\bar{n}_j + 1) \cD[f_j(\ophi)\osigma_{-}]\rho + \kappa \bar{n}_j \cD[f_j(\ophi)\osigma_{+}]\rho \nonumber
\end{eqnarray}
This master equation is valid only in the weak-coupling limit of $\hbar/I,g,\la \oL\ra /I,\kappa \ll \omega_0$, where the bare qubit frequency $\omega_0$ can be removed by switching into the rotating frame \cite{rivas2010,levy2014,hofer2017,gonzalez2017}.

Here, we see two key differences between autonomous and externally-driven engines. Firstly, \eqref{eq:MEauto} describes the evolution of a composite system that comprises both the working medium and the rotor. This means that the cycle period is no longer fixed since the frequency is now a dynamical variable with mean $\mean\oL/I$. Second, the Lindblad operators in the dissipators depend on $\ophi$ and are thus operators acting on the rotor Hilbert space as well. Hence, they will not only describe the energy exchange of the working medium with the hot and cold reservoirs, but also an effective measurement backaction in the form of angular momentum diffusion. This results in an accumulation of passive, disordered energy in the rotor, with its heating rate given by

\begin{eqnarray} \label{eq:thermalDiss}
\dot{\cQ}_{\rm BA} =\tr \left\{\frac{\oL^2}{2I}(\cL_{\rm h}^{\ophi}+\cL_{\rm c}^{\ophi}) \rho \right\}  =\frac{\hbar^2 \kappa}{4I}\sum_{j} \tr \left\{ (\bar{n}_j + |e\ra\la e|) [f_j' (\ophi)]^2 \rho \right\} \geq 0.
\end{eqnarray}
Concerning the engine's energy balance, we note once again that while $\dot{\cQ}_{\rm BA}$ contributes to the total heat flow, it can be omitted in the overall energy balance in the weak-coupling limit where $g,\hbar/I,\kappa\ll\omega_0$,~see \eqref{eq:heatT}. Also, $\dot{\cQ}_{\rm BA}$ should be distinguished from the amount of \emph{useful} work that accumulates in the form of net directed rotation. One should thus be cautious in choosing how to measure the  amount of useful energy generated under autonomous operation with no load attached.

\subsubsection{Autonomous work production}\label{ssec:workdef}

A main advantage of rotor engines is their unambiguous notion of useful energy output: the working medium generates extractable work by producing a gain in net \emph{directed} rotation. This is intuitively clear from the viewpoint of classical motors and piston engines, but does it translate to the quantum regime?

We first look at the direct translation of the conventional work notion based on \eqref{eq:workT} to autonomous engines. The time derivative of the coupling Hamiltonian now turns into the (symmetrized) product of angular velocity and torque acting on the rotor. This leads to a classically inspired, intrinsic work rate definition,
\begin{equation} \label{eq:Wint}
\dot{\cW}_{\rm int} (t) = - \tr \left\{ \rho(t) \frac{\left\{ \oL,  \partial_\varphi \oH_{\rm int} (\ophi) \right\}}{2I} \right\}.
\end{equation}
It describes the rate of change in kinetic energy of the rotor caused solely by the interaction Hamiltonian, i.e.~the force exerted by the working medium. The overall rate of change in kinetic energy is given by \cite{stella2018}
\begin{equation} \label{eq:Wkin}
\dot{\cW}_{\rm kin} (t) =\dot{\cW}_{\rm int}(t) + \dot{\cQ}_{\rm BA}(t).
\end{equation}
For backaction-free engine designs, these two powers are equivalent, which matches a classical setting. However, as $\la \oL^2\ra$ does not encode any directionality, such a gain could also be obtained from pure heating of the rotor, and it is in general not a reliable measure for the production of useful work.

Heuristically, we can alleviate this problem if we measure the kinetic energy associated only to the net directed motion, $\cW_{\rm net} (t) = \la \oL \ra^2/2I$, and its time derivative\footnote{We assume that the hot and cold dissipators do not themselves contribute to a net boost of angular momentum through their angular dependence. In the present case of thermal dissipators \eqref{eq:thermalDiss}, this is ensured for real-valued modulation functions $f_{\rm h,c}$.}
\begin{equation} \label{eq:Wnet}
\dot{\cW}_{\rm net} (t) = \frac{\diff}{\diff t} \frac{\la\oL\ra^2}{2I} = -\tr \left\{ \rho(t) \frac{\oL}{I} \right\} \tr \left\{ \rho(t) \partial_{\varphi} \oH_{\rm int} (\ophi) \right\}.
\end{equation}
This will always be smaller than the gain in kinetic energy, but it may exceed the intrinsic estimate \eqref{eq:Wint} in the presence of backaction, as we shall see later in Figure \ref{fig:pistonLWork}.

Alternatively, we can employ a formal upper bound on the amount of work that an external agent could ideally obtain from the engine motion at a given point time, \emph{ergotropy} \cite{allah2004work,goold2016review}, as discussed also in \chergo. In this case, an agent accessing the kinetic energy stored in the engine rotor could extract at most
\begin{equation}\label{eq:Werg}
\cW_{\rm erg} (t) = \max_{\oU} \tr \left\{ \frac{\oL^2}{2I} \left[\rho_{\rm r} (t) - \oU \rho_{\rm r}(t) \oU^\da \right]  \right\},
\end{equation}
by means of a unitary that brings the reduced rotor state $\rho_{\rm r} (t)$ to a \emph{passive} state\footnote{A passive state is a state whose energy content cannot be reduced further by means of another unitary. This implies that this state must be diagonal in the energy eigenbasis and its eigenvalues must decrease with growing energy.}. Obviously, we can only speak of work \emph{output} from the qubit-rotor system given an actual physical interface or load through which energy is extracted from the engine to perform a task. Hence, $\cW_{\rm erg}$ here describes an axiomatic theoretical upper bound on the extractable work at a given point in time, rather than a real-time work extraction process. For that, we refer the reader to Sect. \ref{ssload}, where we add a dissipative load to ``put the wheels on the ground".

\subsubsection{Autonomous single-qubit piston engine}

\begin{figure}[t]
\includegraphics[width=\textwidth]{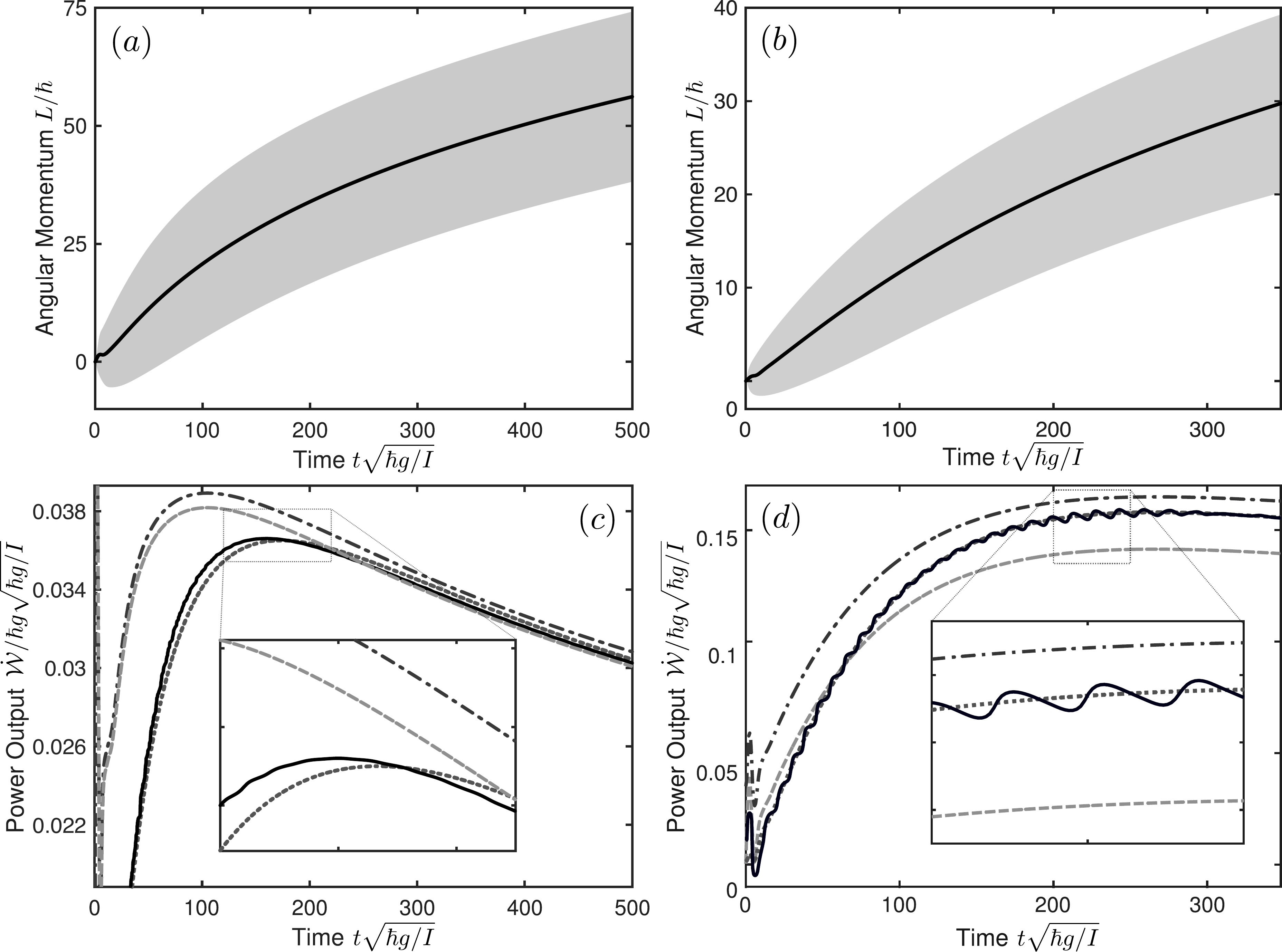}
\caption{Simulation of the engine dynamics for different coupling strength, $g=10\kappa$ (first column) and $g=\kappa$. Panels (a) and (b) depict the time evolution of angular momentum (solid), with the shaded region covering two standard deviations. Panels (c) and (d) compare the different power output definitions: rate of change in ergotropy (solid), kinetic energy (dash-dotted) and net kinetic energy (dotted) with the power associated to the intrinsic torque (dashed). The bath temperatures are $(\bar{n}_{\rm h},\bar{n}_{\rm c})=(1,0.1)$ and we set $I\kappa=10\hbar$.}
\label{fig:pistonLWork}      
\end{figure}

We will now revisit the single-qubit piston engine of Sect.~\ref{sssec:QubitEngineTime} in this autonomous framework. Once again, the simulations are performed on QuTiP, with the rotor Hilbert space truncated to angular momentum eigenstates $|\ell\ra$ with quantum numbers $-60 \leq \ell \leq 200$, and the engine initialized in its ground state $\rho(0)=|e\ra\la e| \otimes |\ell=0\ra\la \ell=0|$. 

Figures \ref{fig:pistonLWork} (a) and (b) compare the time evolution of angular momentum for different coupling strength (a) $g=10\kappa$ and (b) $g=\kappa$, with $(\bar{n}_{\rm h},\bar{n}_{\rm c})=(1,0.1)$ and $I\kappa=10\hbar$. One could show that in the regime where $\la \oL \ra/I \ll \kappa$, ~i.e. when the rotor is just accelerating, there is a linear gain both in the mean angular momentum $\la \oL \ra$ of the rotor (solid line) as well as in its variance (standard deviation shaded)\cite{alex2017rotor}. This means that the initial signal-to-noise ratio (the ratio between the mean and the standard deviation) increases linearly with $\sqrt{t}$. Once the rotor gets into the regime  $\mean\oL/I\sim \kappa$, the acceleration becomes smaller since the qubit could no longer thermalize effectively in this limit, which is similar to having a high driving frequency where $\omega\sim\kappa$ for the externally-driven engine in Sect.~\ref{sssec:QubitEngineTime}. In the plots, this happens at $\mean\oL \approx20\hbar$ where the gain is no longer linear.

Figures \ref{fig:pistonLWork} (c) and (d)  compare the work production rates for the engine motion in (a) and (b) according to the different definitions introduced in Sect.~\ref{ssec:workdef}. We attribute the points of maximum powers to the critical cycle duration that is determined by the finite reaction time, reached when $\mean\oL/I\sim \kappa$. Here, the heating due to backaction $\dot\cQ_\mathrm{BA}$, given by the offset between $\dot{\cW}_{\rm kin}$ (dash-dotted) and $\dot{\cW}_{\rm int}$ (dashed), is greater when $g=\kappa$ (d) as compared to $g=10\kappa$ (c). This is apparent from \eqref{eq:thermalDiss} since $\dot\cW_\mathrm{int}$ scales linearly with $g$ while  $\dot\cQ_\mathrm{BA}$ scales linearly with $\kappa$.

The solid line depicts the time derivative of ergotropy $\dot{\cW}_{\rm erg}$ and is best matched by the rate of change in net kinetic energy $\dot{\cW}_{\rm net}$ (dotted) in both plots, vindicating the intuition derived from using a rotor. For the case of $g=\kappa$ (d), $\dot{\cW}_{\rm erg}$ exhibits an oscillatory pattern around $\dot{\cW}_{\rm net}$. The two curves meet when the mean angular momentum $\mean\oL$ assumes an integer multiple of $\hbar$: this is the consequence of momentum quantization and represents the only genuine quantum signature in the otherwise incoherent rotor dynamics \cite{stella2018}.

\begin{figure}[t]
\sidecaption[t]
\includegraphics[width=7cm]{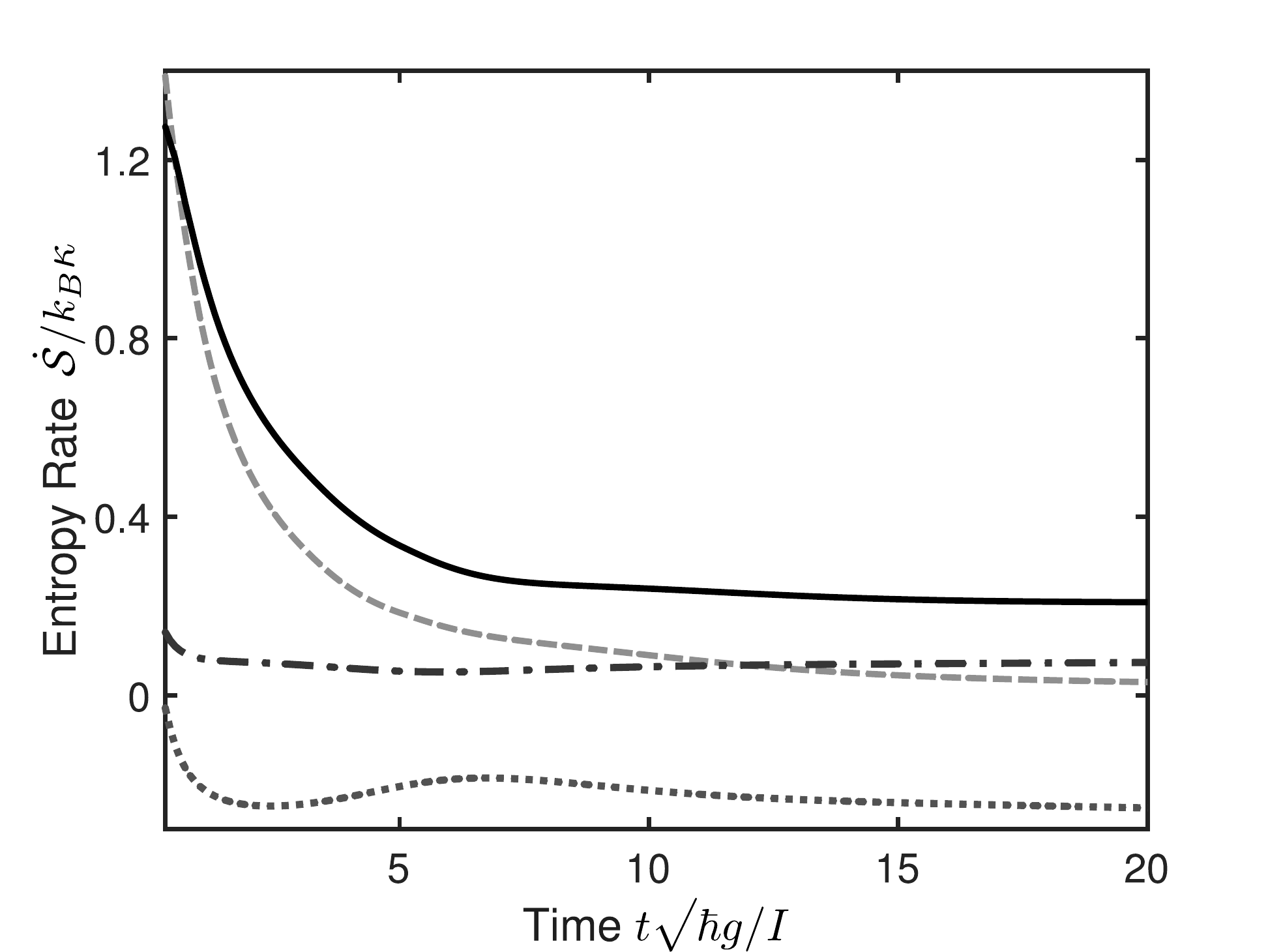}
\caption{Rate of entropy change for the same parameter settings as Figure \ref{fig:pistonLWork}. The net entropy production rate (solid) is obtained by subtracting the entropy rates from the cold (dotted) and hot (dash-dotted) baths from the rate of change in von Neumann entropy of the engine state (dashed).}
\label{fig:pistonEntropy}       
\end{figure}

Leaving the ambiguity of intrinsic work notions aside, the use of \textit{composite} open quantum systems as autonomous engine models also raises the question of thermodynamic consistency. Specifically, the second law may or may not hold, depending on whether and how the master equation describes thermalization of the engine system with each of the reservoirs \cite{alicki1979quantum,levy2014}, see also \chtpi. Here, as a consistency check, we show the relevant entropy rates for $g=10\kappa$ in Figure \ref{fig:pistonEntropy}. The net entropy production rate (solid) is obtained by subtracting the contributions due to the cold (dotted) and hot (dash-dotted) baths, given by $\dot{\cS}_j = \cQ_j/k_BT_j$, from the time derivative of the von Neumann entropy $\cS_\mathrm{sys} = -\tr\left( \rho \ln \rho\right)$ of the engine state (dashed). In this case, we see that the net entropy production would always remain positive as it decays to a steady value of approximately $0.21k_B\kappa$.

\subsubsection{Work extraction by a dissipative load}\label{ssload}

So far, we discussed the transient work output under autonomous operation and illustrated that the kinetic energy associated to the net directed motion of the rotor follows closely with the maximum amount of work that can be extracted (ergotropy). However, we see also the drop in transient powers once the rotor gets into a critical regime $\mean\oL/I\sim \kappa$. Hence, in most practical scenarios, work extraction would instead take place real-time via the attachment of an external load such that the engine operates at a steady state. Here, we introduce a dissipative load by means of an additional Lindblad term $\mathcal{L}_r$ to the master equation \eqref{eq:MEauto}
\begin{equation}\label{eq:Lindbladfriction}
\mathcal{L}_{\rm r}\rho = \frac{2k_{\rm B}T_{\rm r} I\gamma}{\hbar^2} \left( \cD\left[\cos\hat{\varphi} - \frac{i\hbar\sin\hat{\varphi}\oL}{4k_{\rm B}T_{\rm r} I}\right] \rho + \cD \left[\sin\hat{\varphi}+\frac{i\hbar\cos\hat{\varphi}\oL}{4k_{\rm B}T_{\rm r} I}\right] \rho \right) ,
\end{equation}
which describes angular momentum decay at a rate $\gamma$ and effective thermalisation to a Gibbs-like state at temperature $T_r$ \cite{benjamin2017}. This is analogous to the Caldeira-Leggett model of linear Brownian motion \cite{caldeira1983path,Breuer2002}, except that it consists of \emph{two} dissipators expressed in terms of trigonometric functions to maintain $\hat\varphi$-periodicity and rotor symmetry.

\begin{figure}[t]
\includegraphics[width=\textwidth]{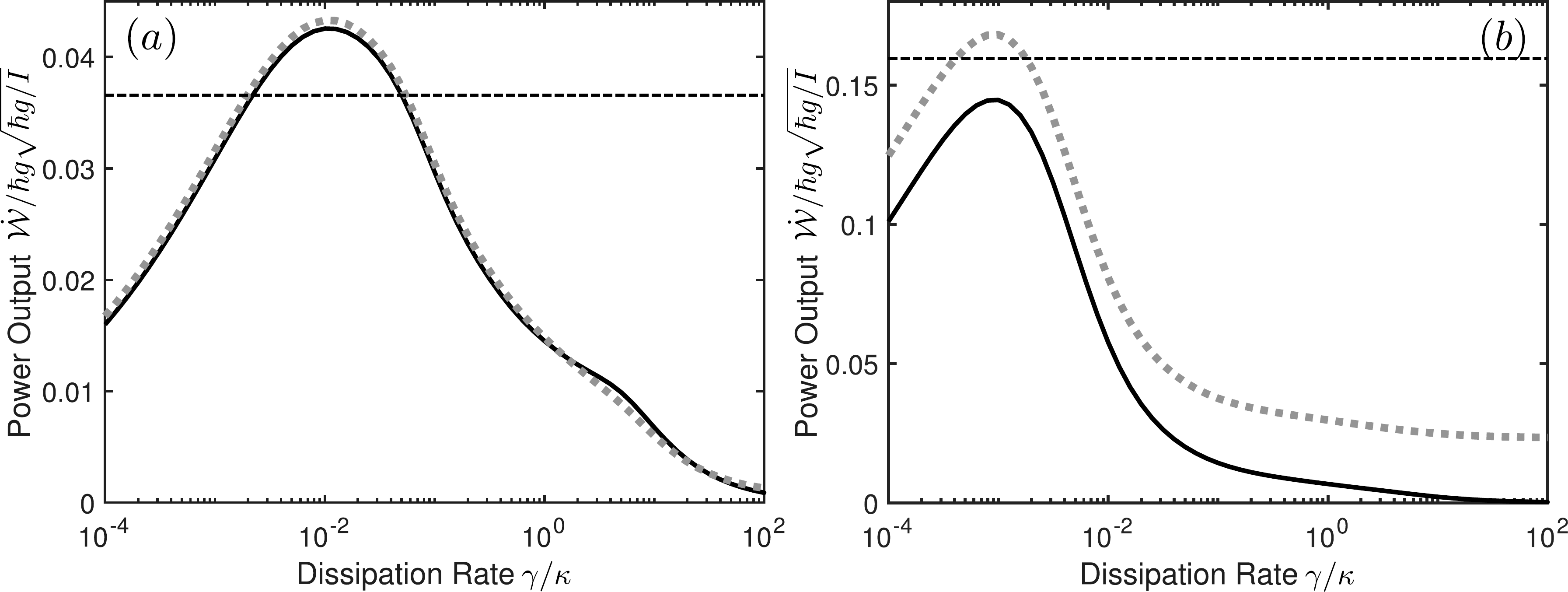}
\caption{Steady-state power associated to the intrinsic torque $\dot\cW_\mathrm{int}$ (solid) and output power to load $\dot\cW_\mathrm{load}$ (dotted) as a function of the dissipation rate $\gamma$ for the same parameters as Figure \ref{fig:pistonLWork}, with (a) $g=10\kappa$ and (b) $g=\kappa$. The horizontal dashed lines  mark the greatest ergotropy rate as depicted in Figure \ref{fig:pistonLWork}. Here, we set $k_B T_r = 10\hbar^2/I$.} 
\label{fig:pistonLoad}       
\end{figure}

At steady-state operation, the output power $\dot\cW_\mathrm{load}$ is given by
\begin{equation}\label{eq:Wload}
\dot{\cW}_\mathrm{load} = - \tr \curly{ \left[ \frac{\oL^2}{2I} + \oH_{\rm int} (\ophi) \right] \cL_{\rm r} \rho } \approx \dot{\cW}_\mathrm{int}+ \dot\cQ_\mathrm{BA},
\end{equation}
and we expect that this balances the net heat flow from the thermal baths. Figure \ref{fig:pistonLoad} shows the steady-state $\dot\cW_\mathrm{int}$ (solid) and $\dot\cW_\mathrm{load}$ (dotted) for different coupling strength $g$ where (a) $g=10\kappa$ and (b) $g=\kappa$, with the same parameter settings as Figure \ref{fig:pistonLWork}. We see that the maximum steady-state power to the load (dotted) exceeds the maximum for the ergotropy rate without load (dashed line). This is because in contrast with ergotropy, a dissipative load would not be able to distinguish between the energy associated to useful directed motion and passive energy and could in principle extract more work in real-time. Notice also that the two output powers deviate more for small coupling $g$ (b) as a result of a greater heating due to backaction, seen also in the transient case (Figure \ref{fig:pistonLWork}). In the limit where $\gamma\rightarrow\infty$, $\dot\cW_\mathrm{int}$ decays to zero, that is the rotor is eventually stopped by the large dissipation rate. In the same limit, $\dot\cW_\mathrm{load}>0$ as a result of backaction-induced momentum diffusion due to continuous exchange of excitations between the baths and the working medium, which enters the rotor via the coupling functions $f(\varphi)$.

\section{Classical vs Quantum}
\label{sec:classical}

Having introduced the model of the engine, we can go back to the main motivation stated in the introduction: look for genuine quantum features that may either enhance or deteriorate the performance of thermal machines, compared to a classical implementation.

One systematic approach to this problem, which is discussed in detail in \chqfs{}, is to identify the effects of quantum coherence in a given quantum system by comparing it to an incoherent stochastic process with equivalent energy balance. This approach is at risk of wiping out any notion of coherence, including the coherence between modes that is allowed in classical physics.

Here, we rather look for a classical description of the degrees of freedom and their dynamics. The two approaches are known to give different results \cite{nimmrichter2017}. Building such a classical analog requires some choices. For instance, in a dissipator like \eqref{eq:thermalDiss}, one has to decide how to deal with the contribution of spontaneous emission. Here, in order to translate the single-qubit piston engine into a classical framework, we need to define a classical system that mimics a qubit. We explore two possibilities: a heuristic coin flip model where the qubit is replaced by a bit, much in the spirit of the coherent-versus-incoherent method; and a more physical spin precession model, where the qubit is turned into a classical magnetic moment.

Further, in the comparison, the initial condition of the quantum evolution of the rotor should no longer be its delocalized ground state of motion $|\ell = 0\ra$, because it has no classical analogue. Instead, we employ a localized wave packet at rest for the initial rotor state, described by the periodic von Mises wavefunction \cite{fisher1995vonmises}
\begin{equation}\label{eq:vm}
\la \varphi |\psi \ra = \frac{e^{\cos\left(\varphi-\mu\right)/2\sigma_{\varphi}^2}}{\sqrt{2\pi I_0\left(\sigma_\varphi^{-2}\right)}},
\end{equation}
with $I_0\left(\sigma_\varphi^{-2}\right)$ a modified Bessel function. This choice approximates a Gaussian wave packet on the circle, localized at rotor position $\mu_\varphi$ with standard deviation $\sigma_\varphi$ in the limit where $\sigma_\varphi\ll 1$. It follows that the momentum distribution is also approximately Gaussian with  standard deviation satisfying $\sigma_\ell ^2 \sigma_\varphi ^2= 1/2$. In classical simulations, where the angle can always be unwrapped to an unbounded coordinate, we can emulate this quantum state easily by considering Gaussian distributions for both the rotor's position and momentum. In all subsequent comparisons, we consider an initial rotor state defined by $(\mu_\varphi,\sigma_\varphi^2) =(\pi/2, 0.1 )$ and $(\mu_\ell,\sigma_\ell^2) =(0, 10)$.



\subsection{Coin model}
\label{ssec:coinflip}

Heuristically, we can recast the autonomous single-qubit engine model introduced in Sect.~\ref{ssec:autoEngine} into a classical system comprised of one bit and a rotor, driven by telegraphic noise with an angle-dependent bias.  For this, we note that the quantum engine dynamics in \eqref{eq:MEauto} depends solely on whether the qubit is excited, but not on its coherence between ground and excited state. Hence, we can simply read off the angle-dependent excitation rate from the master equation and introduce discrete noise that flips a classical ``coin'' bit between its states 0 and 1 at the rate\footnote{Notice that we choose to keep the contribution of spontaneous emission.}
\begin{equation}\label{eq:probrates}
\dot{p}_0 (\varphi) = -\dot{p}_1 (\varphi) = \kappa\sum_{j={\rm h,c}}{ f_{j}^2 (\varphi)\left[(\bar{n}_{j}+1) p_1 (\varphi) - \bar{n}_{j} p_0 (\varphi) \right]}.
\end{equation} When excited, the coin shall exert a torque on the classical rotor, which can be represented by a set of stochastic differential equations for the coin state $C_t \in \{0,1\}$ and the rotor variables,
\begin{eqnarray}\label{eq:coinSDE}
\diff C_t &=& [1-C_t] \diff N_0 - C_t \diff N_1,\\
\diff \varphi_t &=& \frac{L_t }{I}\diff t, \quad \diff L_t = \hbar g C_t \sin \varphi_t \diff t. \nonumber
\end{eqnarray}
The $\diff N_{0,1} \in \{0,1\}$ are two independent random increments with expectation values
\begin{equation}\label{eq:coindN}
\cE \left[ \diff N_{m} \right]_t = \kappa \left[ (\bar{n}_{\rm h}+m) f_{\rm h}^2 (\varphi_t) + (\bar{n}_{\rm c}+m) f_{\rm c}^2 (\varphi_t) \right] \diff t.
\end{equation}

We remark that the flipping noise in the coin model is simply a one-to-one translation of the quantum noise using the same parameters, i.e.~it is not derived from a physical model of energy exchange with a classical thermal bath. This highlights the possibility of using alternative entropy sources apart from standard thermal baths to drive quantum or classical engines.

\begin{figure}[t]
\includegraphics[width=\textwidth]{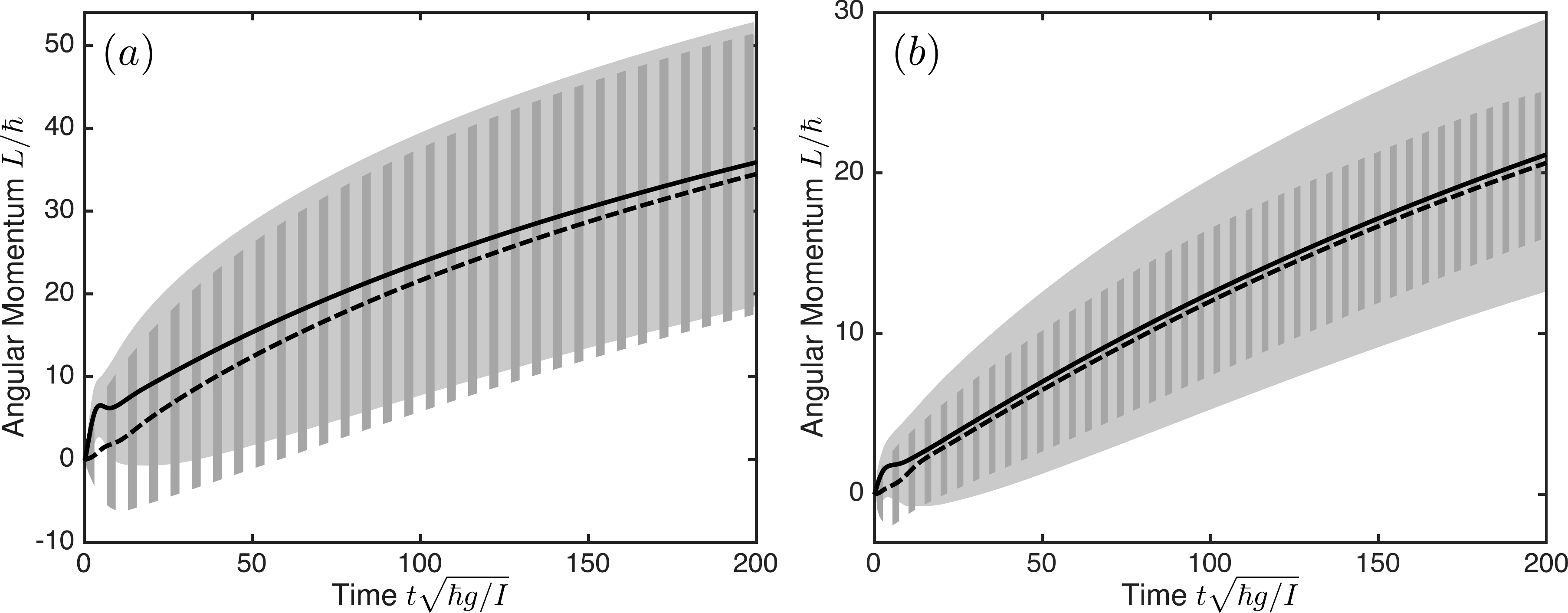}
\caption{Comparison between engine dynamics of quantum (solid) and classical coin flip model (dashed), obtained from averaging over $3.2\times10^7$ trajectories. The lines represent the mean angular momentum while the shaded regions cover two standard deviations.}
\label{fig:QvsCoin}       
\end{figure}

Figure \ref{fig:QvsCoin} compares the engine dynamics for different coupling strength $g$, similar to the settings used in Figure \ref{fig:pistonLoad} where (a)  $g=10\kappa$ and (b) $g=\kappa$. For both parameter settings, the coin flip model (dashed) predicts similar average behavior but the difference in terms of momentum noise is apparent for small $g$. This is because contrary to the quantum case, the coin model is backaction-free, i.e.~the noise input that flips the coin does not affect the increment of the angular momentum variable. On the other hand, the Lindblad dissipators \eqref{eq:thermalDiss} in the quantum model describe momentum diffusion or, complementarily, \emph{decoherence} in the angle representation and we see earlier from Figure \ref{fig:pistonLoad} that backaction noise is more prominent for small $g$.


\subsection{Magnetic moment model}
We have shown that the coin flip model, while reproducing the average behavior of the quantum engine, is unable to capture the effects of backaction noise. To include this aspect of the quantum dynamics in a classical setting, we consider a physical model describing a precessing magnetic moment\cite{palacios2007classicspin}, which now replaces the qubit as the working medium.

In this case, a classical magnetic moment vector $\mu_0 \vec{m}$ precesses about an external magnetic field $\vec{B} = B_\mathrm{ext}\vec{\hat{z}}$, where $\vec{m}=m_x \vec{\hat x}+m_y \vec{\hat y}+m_z \vec{\hat z}$  can be interpreted as the analogue of a quantum spin vector. The free Hamiltonian of the working medium is then given by $H_s = \hbar \omega_0 m_z$, where $\omega_0 = \mu_0 B_\mathrm{ext} /\hbar$. The interaction between the rotor and the spin is determined by the magnitude of magnetic moment along $+\vec{\hat{z}}$, and is described by $H_\mathrm{int} =\hbar g (m+m_z) \cos\varphi$. This system is then coupled linearly to a set of classical harmonic oscillators that serves as the thermal baths via the coupling functions $f_j(\varphi)$ introduced earlier. The Langevin equations governing the dynamics of the engine are given by
\begin{eqnarray}\label{eq:classical}
\mathrm{d}m_z = -\kappa\left[f_{\rm h}^2(\varphi)+f_{\rm c}^2(\varphi)\right] \frac{m^2 - m_z^2}{m} \mathrm{d}t - 2\kappa\left[\epsilon_{\rm h} f_{\rm h}^2(\varphi)+\epsilon_{\rm c} f_{\rm c}^2(\varphi)\right]\frac{m_z}{m}\mathrm{d}t\nonumber\\+\sqrt{2\kappa\left(\epsilon_{\rm h} f_{\rm h}^2(\varphi)+\epsilon_{\rm c} f_{\rm c}^2(\varphi)\right)\frac{m^2 - m_z^2}{m}}\mathrm{d}W_1, \nonumber\\
\mathrm{d}L = \hbar g(m+m_z) \sin \varphi\mathrm{d}t +\hbar\sqrt{2\kappa\left(\epsilon_{\rm h} f_{\rm h}'^2(\varphi)+\epsilon_{\rm c} f_{\rm c}'^2(\varphi)\right)\frac{m^2 - m_z^2}{m}}\mathrm{d}W_2,
\end{eqnarray}
where $\diff W_{1,2}$ denote two independent real-valued Wiener increments of variance $\diff t$. 

Note that backaction enters this model in the noise term present in the equation of motion for the angular momentum. For a physical comparison between the two models, we assume the same bath temperatures, where the classical excitation numbers are now determined by $\epsilon_\mathrm{h,c} = k_B T_\mathrm{h,c}/ \hbar\omega_0$. We also set the magnitude of the spin vector to $m = 1/2$. 

\begin{figure}[t]
\includegraphics[width=\textwidth]{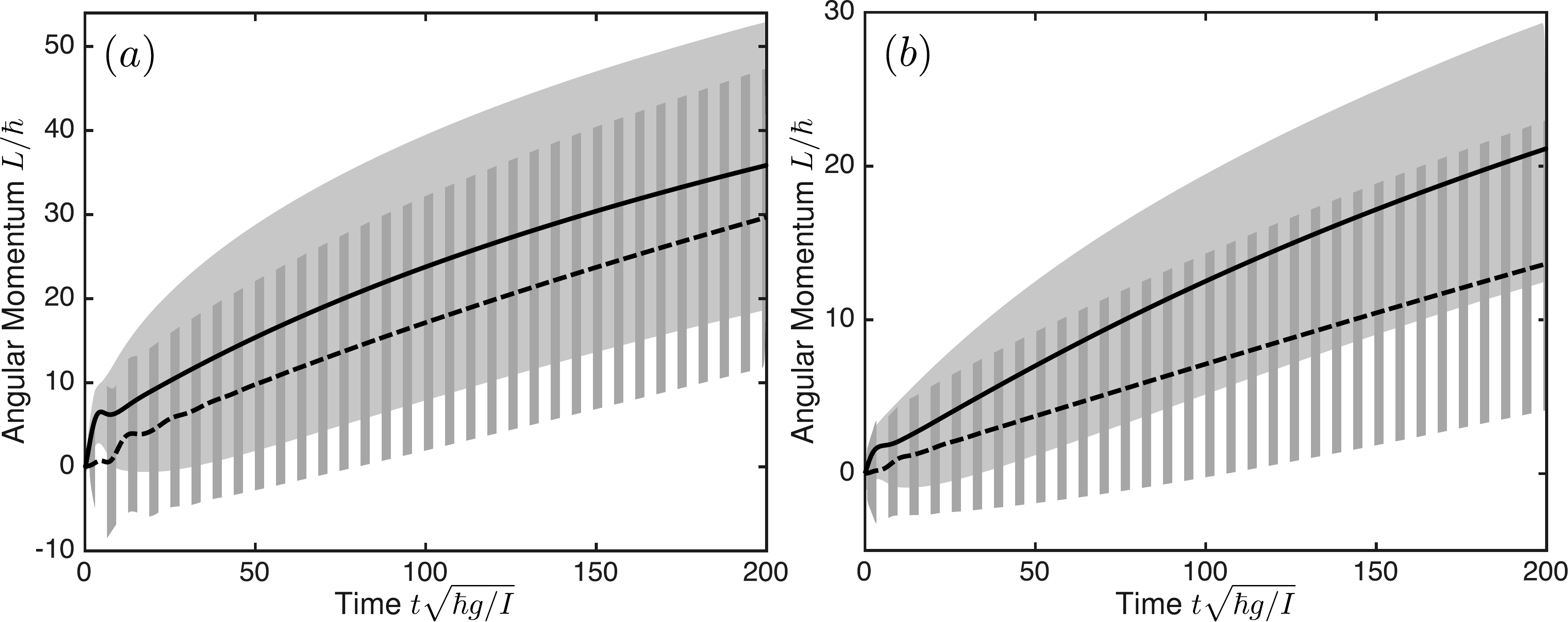}
\caption{Comparison between engine dynamics of quantum (solid) and classical magnetic moment model (dashed), obtained from averaging over $3.2\times10^7$ trajectories. The lines represent the mean angular momentum while the shaded regions cover two standard deviations.}
\label{fig:QvsMagnMom}       
\end{figure}

Similar to previous settings, Figure \ref{fig:QvsMagnMom} compares the quantum and classical engine dynamics at two different coupling strength. As expected, the classical spin model (dashed) predicts greater momentum noise as a result of the backaction term, which is now comparable to the quantum prediction (solid) both in the case of large (a) and small (b) coupling strength. However, the average behavior differs significantly, which we attribute to the different thermal statistics. When considering a spin-$1/2$ of frequency $\omega_0$ coupled to a bath of temperature $T$, we get
\begin{eqnarray}\label{eq:meanm}
\mean{m_z} = \frac{k_BT}{\hbar\omega_0}-\frac{1}{2}\coth\left(\frac{\hbar\omega_0}{2k_B T}\right),\quad
\mean{\osigma_z} = -\frac{1}{2} \tanh\left(\frac{\hbar\omega_0}{2k_B T}\right),
\end{eqnarray}    
since the classical spin assumes a continuous $m_z$ value ranging from $\left[-1/2,1/2\right]$ while the quantum one only takes discrete values $\pm 1/2 $. 
In principle, one could align the average behaviors by matching the mean spin values, but that would imply different bath temperatures for the quantum and the classical case.

\section{Conclusion}
Drawing analogy from classical engines capable of generating directed motion, we introduced the use of rotors in quantum engines. We first started off with a textbook example of a driven engine cycle, and demonstrated how the use of a rotor facilitates the mapping of the time-dependence $\omega t$ to an autonomous setting characterized by $\left(I,\ophi\right)$ and see explicitly the role of the rotor as an internal clock that sets the engine cycle as well as a work storage. In particular, there is an intuitive notion of the \emph{useful} work stored, in that the energy associated to the net motion actually matches the axiomatic maximum work extraction (ergotropy). We then looked at real-time work extraction by subjecting the rotor to an external dissipative load, and showed that the steady-state power can exceed the maximum ergotropy rate in the transient operation. 

Finally, we compared the dynamics of the engine in entirely classical frameworks. If the qubit is replaced with a coin-flip model driven by telegraphic noise, the predicted dynamics is less noisy because one can’t introduce backaction noise. If the qubit is replaced by a classical magnetic moment driven by linearly coupled harmonic oscillators, the average behavior is somewhat worse for a given choice of the temperatures of the baths. While these differences, at first glance, may point to quantum effects (a detrimental one in the first case and an advantageous one in the second), the general behaviors are similar to the quantum one: in particular, we do not find any evidence of ``quantum supremacy''. However, rotor-based engines with classical counterparts provide a testbed for future studies of genuine quantum effects in quantum heat engines.

\begin{acknowledgement}
This research is supported by the Singapore Ministry of Education through the Academic Research Fund Tier 3 (Grant No. MOE2012-T3-1-009); and by the same MoE and the National Research Foundation, Prime Minister's Office, Singapore, under the Research Centres of Excellence programme. In addition, this work was financially supported by the Swiss SNF and the NCCR Quantum
Science and Technology.
\end{acknowledgement}
%

%


\begin{thebibliography}{10}
\providecommand{\url}[1]{{#1}}
\providecommand{\urlprefix}{URL }
\expandafter\ifx\csname urlstyle\endcsname\relax
  \providecommand{\doi}[1]{DOI \discretionary{}{}{}#1}\else
  \providecommand{\doi}{DOI \discretionary{}{}{}\begingroup
  \urlstyle{rm}\Url}\fi

\bibitem{alex2017rotor}
A.~Roulet, S.~Nimmrichter, J.M. Arrazola, S.~Seah, V.~Scarani, Phys. Rev. E
  \textbf{95}, 062131 (2017).
\newblock \doi{10.1103/PhysRevE.95.062131}.
\newblock \urlprefix\url{https://link.aps.org/doi/10.1103/PhysRevE.95.062131}

\bibitem{stella2018}
S.~Seah, S.~Nimmrichter, V.~Scarani, New Journal of Physics \textbf{20}(4),
  043045 (2018).
\newblock \urlprefix\url{http://stacks.iop.org/1367-2630/20/i=4/a=043045}

\bibitem{alex2018}
A.~Roulet, S.~Nimmrichter, J.M. Taylor, arXiv:1802.05486  (2018).
\newblock \urlprefix\url{https://arxiv.org/abs/1802.05486}

\bibitem{tonner2005}
F.~Tonner, G.~Mahler, Phys. Rev. E \textbf{72}, 066118 (2005).
\newblock \doi{10.1103/PhysRevE.72.066118}.
\newblock \urlprefix\url{https://link.aps.org/doi/10.1103/PhysRevE.72.066118}

\bibitem{youssef2010}
M.~Youssef, G.~Mahler, A.S. Obada, Physica E \textbf{42}(3), 454  (2010).
\newblock \doi{https://doi.org/10.1016/j.physe.2009.06.032}.
\newblock
  \urlprefix\url{http://www.sciencedirect.com/science/article/pii/S1386947709002355}

\bibitem{brunner2012virtual}
N.~Brunner, N.~Linden, S.~Popescu, P.~Skrzypczyk, Phys. Rev. E \textbf{85},
  051117 (2012).
\newblock \doi{10.1103/PhysRevE.85.051117}.
\newblock \urlprefix\url{https://link.aps.org/doi/10.1103/PhysRevE.85.051117}

\bibitem{gilz2013}
L.~Gilz, E.P. Thesing, J.R. Anglin, arXiv preprint arXiv:1304.3222  (2013).
\newblock \urlprefix\url{https://arxiv.org/abs/1304.3222}

\bibitem{mari2015quantum}
A.~Mari, A.~Farace, V.~Giovannetti, J. Phys. B \textbf{48}(17), 175501 (2015).
\newblock \urlprefix\url{http://stacks.iop.org/0953-4075/48/i=17/a=175501}

\bibitem{kosloff2016flywheel}
A.~Levy, L.~Di\'osi, R.~Kosloff, Phys. Rev. A \textbf{93}, 052119 (2016).
\newblock \doi{10.1103/PhysRevA.93.052119}.
\newblock \urlprefix\url{https://link.aps.org/doi/10.1103/PhysRevA.93.052119}

\bibitem{hardal2017}
A.U.C. Hardal, N.~Aslan, C.M. Wilson, O.E. M\"ustecapl\ifmmode \imath \else \i
  \fi{}o\ifmmode~\breve{g}\else \u{g}\fi{}lu, Phys. Rev. E \textbf{96}, 062120
  (2017).
\newblock \doi{10.1103/PhysRevE.96.062120}.
\newblock \urlprefix\url{https://link.aps.org/doi/10.1103/PhysRevE.96.062120}

\bibitem{kosloff1984}
R.~Kosloff, J. Chem. Phys. \textbf{80}(4), 1625 (1984).
\newblock \doi{10.1063/1.446862}

\bibitem{scully2003}
M.O. Scully, M.S. Zubairy, G.S. Agarwal, H.~Walther, Science
  \textbf{299}(5608), 862 (2003).
\newblock \doi{10.1126/science.1078955}.
\newblock \urlprefix\url{http://science.sciencemag.org/content/299/5608/862}

\bibitem{rezek2006}
Y.~Rezek, R.~Kosloff, New J. Phys. \textbf{8}(5), 83 (2006).
\newblock \urlprefix\url{http://stacks.iop.org/1367-2630/8/i=5/a=083}

\bibitem{alicki2014}
R.~Alicki, Open Syst. Inf. Dyn. \textbf{21}(01n02), 1440002 (2014).
\newblock \doi{10.1142/S1230161214400022}

\bibitem{zhang2014}
K.~Zhang, F.~Bariani, P.~Meystre, Phys. Rev. Lett. \textbf{112}, 150602 (2014).
\newblock \doi{10.1103/PhysRevLett.112.150602}.
\newblock
  \urlprefix\url{http://link.aps.org/doi/10.1103/PhysRevLett.112.150602}

\bibitem{uzdin2016}
R.~Uzdin, A.~Levy, R.~Kosloff, Entropy \textbf{18}(4) (2016).
\newblock \doi{10.3390/e18040124}.
\newblock \urlprefix\url{http://dx.doi.org/10.3390/e18040124}

\bibitem{alicki1979quantum}
R.~Alicki, J.~Phys.~A: Math.~Gen. \textbf{12}(5), L103 (1979).
\newblock \doi{10.1088/0305-4470/12/5/007}

\bibitem{huang2012}
X.L. Huang, T.~Wang, X.X. Yi, Phys. Rev. E \textbf{86}, 051105 (2012).
\newblock \doi{10.1103/PhysRevE.86.051105}.
\newblock \urlprefix\url{https://link.aps.org/doi/10.1103/PhysRevE.86.051105}

\bibitem{obinana2014}
O.~Abah, E.~Lutz, EPL (Europhysics Letters) \textbf{106}(2), 20001 (2014).
\newblock \urlprefix\url{http://stacks.iop.org/0295-5075/106/i=2/a=20001}

\bibitem{rossnagel2014}
J.~Ro\ss{}nagel, O.~Abah, F.~Schmidt-Kaler, K.~Singer, E.~Lutz, Phys. Rev.
  Lett. \textbf{112}, 030602 (2014).
\newblock \doi{10.1103/PhysRevLett.112.030602}.
\newblock
  \urlprefix\url{http://link.aps.org/doi/10.1103/PhysRevLett.112.030602}

\bibitem{niedenzu2016}
W.~Niedenzu, D.~Gelbwaser-Klimovsky, A.G. Kofman, G.~Kurizki, New Journal of
  Physics \textbf{18}(8), 083012 (2016).
\newblock \urlprefix\url{http://stacks.iop.org/1367-2630/18/i=8/a=083012}

\bibitem{klaers2017}
J.~Klaers, S.~Faelt, A.~Imamoglu, E.~Togan, Phys. Rev. X \textbf{7}, 031044
  (2017).
\newblock \doi{10.1103/PhysRevX.7.031044}.
\newblock \urlprefix\url{https://link.aps.org/doi/10.1103/PhysRevX.7.031044}

\bibitem{wulfert2017}
R.~Wulfert, M.~Oechsle, T.~Speck, U.~Seifert, Phys. Rev. E \textbf{95}, 050103
  (2017).
\newblock \doi{10.1103/PhysRevE.95.050103}.
\newblock \urlprefix\url{https://link.aps.org/doi/10.1103/PhysRevE.95.050103}

\bibitem{rivas2010}
A.~Rivas, A.D.K. Plato, S.F. Huelga, M.B. Plenio, New J. Phys. \textbf{12}(11),
  113032 (2010).
\newblock \doi{10.1088/1367-2630/12/11/113032}

\bibitem{levy2014}
A.~Levy, R.~Kosloff, Europhys. Lett. \textbf{107}(2), 20004 (2014).
\newblock \doi{10.1209/0295-5075/107/20004}.
\newblock \urlprefix\url{https://doi.org/10.1209/0295-5075/107/20004}

\bibitem{hofer2017}
P.P. Hofer, M.~Perarnau-Llobet, L.D.M. Miranda, G.~Haack, R.~Silva, J.B. Brask,
  N.~Brunner, New J. Phys. p. in press (2017).
\newblock \doi{10.1088/1367-2630/aa964f}

\bibitem{gonzalez2017}
J.O. Gonz{\'a}lez, L.A. Correa, G.~Nocerino, J.P. Palao, D.~Alonso, G.~Adesso,
  arXiv preprint arXiv:1707.09228  (2017).
\newblock \urlprefix\url{https://arxiv.org/abs/1707.09228}

\bibitem{johansson2013qutip}
J.~Johansson, P.~Nation, F.~Nori, Comput. Phys. Commun. \textbf{184}(4), 1234
  (2013)

\bibitem{allah2004work}
A.E. Allahverdyan, R.~Balian, T.M. Nieuwenhuizen, Europhys. Lett.
  \textbf{67}(4), 565 (2004).
\newblock \urlprefix\url{http://stacks.iop.org/0295-5075/67/i=4/a=565}

\bibitem{goold2016review}
J.~Goold, M.~Huber, A.~Riera, L.~del Rio, P.~Skrzypczyk, J. Phys. A
  \textbf{49}(14), 143001 (2016).
\newblock \urlprefix\url{http://stacks.iop.org/1751-8121/49/i=14/a=143001}

\bibitem{benjamin2017}
B.A. Stickler, B.~Schrinski, K.~Hornberger, arXiv preprint arXiv:1712.05163
  (2017).
\newblock \urlprefix\url{https://arxiv.org/abs/1712.05163}

\bibitem{caldeira1983path}
A.~Caldeira, A.~Leggett, Physica A \textbf{121}(3), 587  (1983).
\newblock \doi{https://doi.org/10.1016/0378-4371(83)90013-4}.
\newblock
  \urlprefix\url{http://www.sciencedirect.com/science/article/pii/0378437183900134}

\bibitem{Breuer2002}
H.P. Breuer, F.~Petruccione, \emph{{The Theory of Open Quantum Systems}}
  (Oxford University Press, 2002).
\newblock
  \urlprefix\url{http://books.google.com/books?id=0Yx5VzaMYm8C{\&}pgis=1}

\bibitem{nimmrichter2017}
S.~Nimmrichter, J.~Dai, A.~Roulet, V.~Scarani, {Quantum} \textbf{1}, 37 (2017).
\newblock \doi{10.22331/q-2017-12-11-37}.
\newblock \urlprefix\url{https://doi.org/10.22331/q-2017-12-11-37}

\bibitem{fisher1995vonmises}
N.I. Fisher, \emph{Statistical analysis of circular data} (Cambridge University
  Press, 1995)

\bibitem{palacios2007classicspin}
J.L. Garc\'{i}a-Palacios, \emph{On the Statics and Dynamics of
  Magnetoanisotropic Nanoparticles} (Wiley-Blackwell, 2007), pp. 1--210.
\newblock \doi{10.1002/9780470141717.ch1}.
\newblock
  \urlprefix\url{https://onlinelibrary.wiley.com/doi/abs/10.1002/9780470141717.ch1}

\end{thebibliography}
\end{document}